\documentstyle[floats,multicol,aps,epsf,prl]{revtex}
\begin{document}
\draft
% ***********    This is for two columns
% *******************************
\twocolumn[\hsize\textwidth\columnwidth\hsize\csname
@twocolumnfalse\endcsname
%********************************
%\documentstyle[eqsecnum,aps]{revtex}
\def\btt#1{{\tt$\backslash$#1}}
\title{ Equilibration and Dynamic Phase Transitions of a Driven Vortex
Lattice}
\author{ Z. L. Xiao and  E. Y. Andrei}
\address{Department of Physics and Astronomy, Rutgers University,
Piscataway, New Jersey 08854}
\author{P. Shuk and M. Greenblatt}
\address{Department of Chemistry, Rutgers University, Piscataway, New
Jersey 08854}
\maketitle
\begin{abstract}
{  We report on the observation of  two  types of current driven transitions in 
metastable vortex lattices. The metastable 
states, which are missed in usual slow transport measurements, are
detected with  a fast 
transport technique in the vortex
lattice of undoped 
 2H-NbSe$_2$.  The  transitions are seen by following the evolution of
these states 
 when driven by a current.   At low currents we observe an 
equilibration 
transition from a metastable to a stable  state, followed by a dynamic
crystallization transition  at high currents.} 

\end{abstract}
%\baselineskip=24 pt
%\draft
]
%\narrowtext  

Vortices in type II superconductors are easily trapped in long lived metastable 
states created 
by  the random pinning potential\cite{blat}.  A driving  current 
 assists in vortex de-trapping and in finding the stable 
states\cite{thor,word,hend,xiao},  in essence assuming the role of 
temperature in ordinary phase transitions. This analogy was pointed out by 
Koshelev-Vinokur (KV) who showed that when the vortices are set in motion  
by a current  $I$, the random potential appears as a temporally
fluctuating Langevin force in the moving frame \cite{kosh}. The random potential 
can then be replaced by a  "shaking temperature" $T_s \propto 1/I $, which leads 
to  a simpler problem of a pure system  at an effective temperature 
$T_{eff}=T+T_s$.  Thus,  if  the pure system crystallizes at a temperature 
$T_m$, the actual  system  will crystallize when driven by a current $I_t\propto 
(T_m-T)^{-1}$.  This KV
transition separates a disordered state  at low currents from an ordered one at 
high currents. Further  work \cite{shi,moon,giam,ols,bal,kolt}  
 predicted that the transition is preceded by a 
regime of plastic flow followed by smectic ordering. Small angle neutron 
scattering (SANS) on the field-cooled (FC) vortex lattice in  
undoped 2H-NbSe$_2$ \cite{yar} provided evidence for current induced ordering. 
This data are  generally 
accepted as direct evidence for the KV transition\cite{giam,bal,sans} despite 
the fact that for usual slow transport measurements under the same experimental 
conditions the transition is absent \cite{yar,bhat,com}. This discrepancy could 
be an indication of metastability in the  
FC vortex state, in which case the observed ordering in the SANS data  would be 
an equilibration rather than the KV transition. Thus far however metastability 
in the vortex lattice of undoped 2H-NbSe$_2$ was thought to be absent 
\cite{yar}. 

In this letter we present results of fast transport measurements that 
demonstrate the existence of metastable states in undoped 2H-NbSe$_2$. The 
experiments  probe the 
dynamic  transitions with emphasis on distinguishing between equilibration 
and the  KV-transition. We show that the current induced transition in the FC 
state is in fact an  equilibration from a disordered metastable state 
to an ordered stable state and not the KV 
transition. We further find that in the peak effect region (a region  of the 
phase diagram, just below H$_{c2}$ where the  critical current increases with 
field and temperature\cite{word}) the vortex lattice can undergo two current 
induced transitions. The first is an equilibration 
transition observed at low currents which is then  followed by the  KV 
crystallization at much higher currents.

The sample was an  undoped single crystal 2H-NbSe$_2 $ platelet of
dimensions 
 1.5x0.65x0.025mm$^3$. Its critical temperature  $ T_c$ was $ 7.1 $K , 
 the transition width  80 mK and the   normal resistance 
near $T_c$ was $21$x10$ ^{-3} \Omega $. A four probe measurement  
with low resistance AgIn solder contacts was used  to monitor vortex
response. 
The response to fast current ramps - 200A/s- was detected with a fast
(2$\mu s$ response time) amplifier 
 while  the  slow ramp -$5\times 10^{-5}$A/s - 
measurements  were obtained with a 
Keithley 181 nanovoltmeter.    The magnetic field was kept along the $c$ axis 
of the 
sample and the current  was in the $ a-b$ plane. The zero-field-cooled
(ZFC) and the FC vortex lattices were prepared  by applying the
magnetic field after or before
 cooling the sample through $T_c$ respectively in the absence of applied 
current. 
   The degree of order of the vortex lattice was
inferred  from the critical current,   $I_c$, (defined by a  5$\mu $V response 
criterion) 
 by using  the 
Larkin-Ovchinnikov model \cite{LO}, according to which $I_c$ grows with the 
degree of disorder.

In Fig.1 we compare the current-voltage ($I-V$) curves of the FC and ZFC 
lattices. 
When probed with  slow current ramps the response of the two states is identical
and no evidence of metastability is observed. This is  in contrast to results in 
Fe doped 2H-NbSe$_2$ where strong metastability and hysteresis 
were found in slow transport measurements\cite{hend}. Here the 
temperature  dependence of the critical currents (inset of Fig. 1(a)) is the 
same for both states,  and exhibits a 
 pronounced peak effect just below $T_c$. But in spite of the identical
response in slow measurements, 
 the  initial vortex states prepared by ZFC 
and by FC are not identical. 
\begin{figure}[btp]
\epsfxsize=3.8 in
\epsfbox{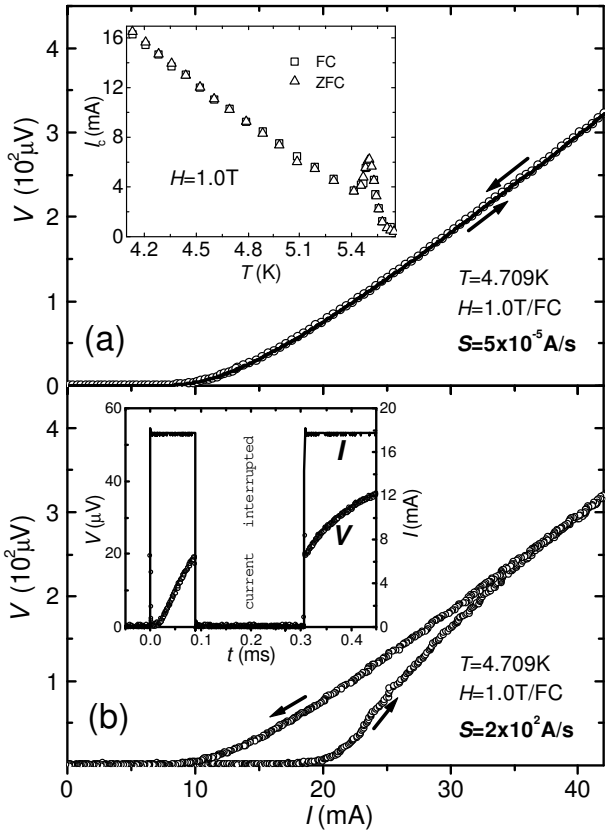} 
\protect\caption{ (a) $I-V$ curves obtained with slow (5x10$^{-5}$A/s) current 
ramps. 
The inset  shows the temperature dependence of the  critical current
for  ZFC and FC vortex lattices obtained with slow measurements. (b)
Fast $I-V$ curves (200A/s) for 
an FC vortex lattice.
  The inset  shows the time evolution of the 
response of an FC vortex lattice to a current step. }
\label{fig:fig_pl1}
\end{figure}
  The difference between the two states becomes
evident  only in   fast   measurements. This is illustrated  in Fig.1
(b) where significant hysteresis is seen when comparing  the  fast $I-V$ curve 
for the pristine  FC  state,
measured  on the first ramp-up of the current, with   that recorded when the 
current is ramped down. The critical
current of the pristine FC state is  almost  twice that of the annealed state
recorded on the down ramp, indicating that it is more disordered. 
By contrast the pristine ZFC state starts out with a low
critical current, is  unchanged by current cycling and its $I-V$ curves are 
independent of ramping speeds.  We conclude that the  ZFC state is ordered and 
stable. For the FC state, the $I-V$ curves obtained after the first ramp up 
(including the ramp down and all subsequent ramps) are identical to those of the 
ZFC state indicating that the FC lattice reorders during the first ramp. It 
follows  that the FC lattice  is initially in a disordered 
metastable state which reorganizes under the influence of a current into a 
stable ordered configuration. 
 
The current driven organization of the FC lattice is 
seen directly in the inset of Fig.1(b) through  the evolution of the  response 
to a current step. After an initial waiting time the response starts growing 
from zero as the vortices order into a state with lower critical current,  and
saturates to  a value that depends on the amplitude of the applied
current. In order to study  the current dependence of the reorganization the 
pristine FC state was driven with long (100sec) current steps  and then quenched 
by suddenly removing the current. When the current is removed the evolution 
of the vortex state is arrested instantaneously and then resumes from the exact 
same 
state when the current is turned back on. This  is clearly seen in the 
inset of Fig.1(b). Thus, 
when  the quenched state is probed by recording the  $I-V$ curve with a fast 
current ramp the critical current reflects the degree of order of the state at 
the moment of quenching. 
\begin{figure}[btp]
\epsfxsize=3.8in
\epsfbox{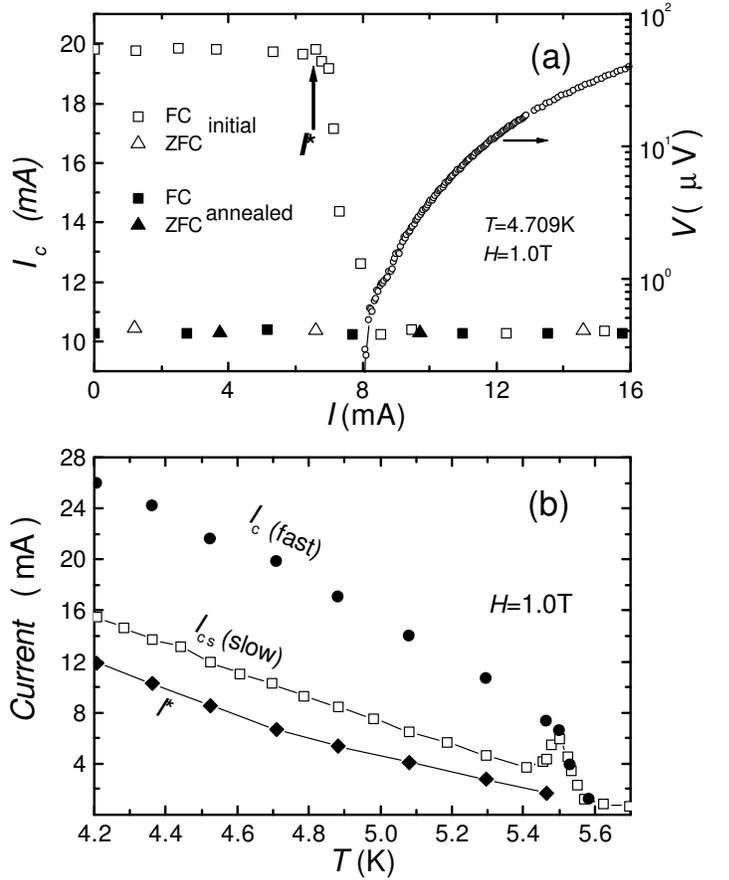} 
\protect\caption{(a) Critical currents for initial and 
annealed vortex lattices prepared by ZFC and FC processes following a
100s drive 
with various currents, $I$. The open circles are the $I-V$ curve of an FC
vortex lattice 
obtained in a slow ($5\times 10^{-5}$A/s) measurement. (b) Temperature 
dependence of $I^*$ and the  critical 
currents of the FC lattice for fast and slow measurements.}
\label{fig:fig_pl1}
\end{figure}
In Fig.2(a) we plot  the critical current  of the quenched state as  a
function of current-step  amplitude. No evolution is seen until $I^*=$6.7mA 
where we note the onset  of a  sharp transition from disordered
state (higher $I_c$) to ordered state (lower $I_c$). The transition is completed 
at $I_{cs}=$8 mA, at which point the voltage response becomes measurable in the 
$I-V$ curve.  Also shown  are  the results for the same experiment on the  
pristine ZFC lattice  and the  annealed  FC and ZFC
lattices. The vortex lattice was  annealed with a 
slow ($5\times 10^{-5}$A/s) cycle of the current between 0 and 50 mA.
The absence of a jump in $I_c$ 
indicates that  the  current does not cause ordering of the  ZFC and annealed
states. 
For  higher amplitude current steps, $I>I_{cs}$  the  $I-V$ curves of all
quenched states  are identical, 
regardless of the initial state.  The transition at $I^*$ 
is the  same as that  seen  in the SANS measurements and interpreted as  the
KV crystallization \cite{yar}. 
But the fact that  it occurs at such low currents 
makes it an unlikely candidate for  the KV transition.  This is 
confirmed by 
 the temperature dependence of $I^*$, shown in  Fig.2(b). Here    $I^*$
decreases 
with increasing temperature, in contrast to the predicted increase with
temperature 
for the  KV-transition.  
We conclude that $I^*$ is not where the KV transition occurs but rather the 
onset of a current 
driven equilibration from the metastable disordered FC state
to a stable ordered state. 
 
\begin{figure}[btp]
\epsfxsize=3.4in
\epsfbox{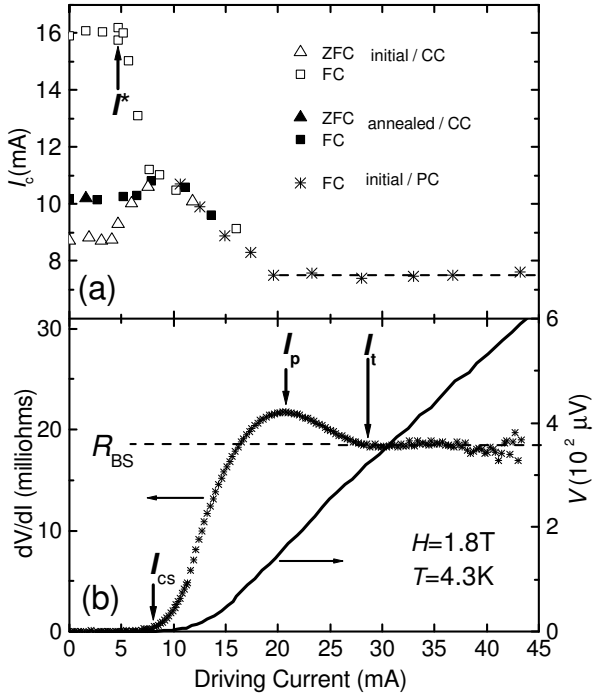} 
\protect\caption{(a) Evolution of the critical current with  driving 
current amplitude for ZFC, FC  and annealed vortex lattices in the peak effect 
region following a 100s drive with 
 continuous (CC) and pulsed (PC) currents. (b) Current dependence 
of
differential resistance 
 illustrating the definition of $I_p$ and $I_t$. The data saturates at 
$R_{BS}$, the Bardeen-Stephen free flux flow value. Also shown is the  $I-V$ 
curve obtained with a pulsed  current.}\label{fig:fig_pl1}
\end{figure}

To further illustrate the difference between the equilibration and the KV 
transition we repeated  the  above experiment in the peak effect region, 
where the KV transition \cite{bhat,kap} and a current driven reorganization 
 in the the ZFC lattice\cite{xiao}  were seen with 
transport measurements. Since much higher currents are 
required in order to 
observe the KV transition, Joule heating needs to be considered. For 
example, for a 45 mA  (the highset current in Fig.3) continuous current drive, a 
temperature increase of 25 mK in the sample was detected. The details on the 
determination of Joule heating in these experiments is  presented elsewhere 
\cite{heat}. To avoid  heating at high currents we apply the current step in 
a sequence of short, 10$\mu$s, pulses separated by 500$\mu$s cooling intervals 
with no current. For currents below 15mA pulsing was not necessary since heating 
was negligible (less than 3mK) and the measured $I_c$ was the same for 
continuous and pulsed applications. In the 
following discussion we focus on the data which show no heating. In Fig.3(a) we 
show the critical current of the quenched vortex lattice following the 
application of 100s current steps of various amplitudes. 
At low currents $I\leq I^*$=4.5mA no
measurable change occurs in the vortex states. For higher currents,  
$I^*<I<I_{cs}$=8.6mA the current driven organization sets in, affecting each
state differently. The critical 
current of the FC state drops rapidly,  that of the ZFC state increases
while the annealed lattice curve is almost unchanged. At
$I_{cs}$ all the curves converge, overlapping at higher currents, which 
indicates that in this regime the vortex state is determined by the driving
current alone and is independent of the initial preparation. The  critical 
current decreases with increasing current step amplitude saturating at a value 
which is lower than that of the stationary vortex lattices, as expected for the 
KV transition.  

\begin{figure}[btp]
\epsfxsize=3.5in
\epsfbox{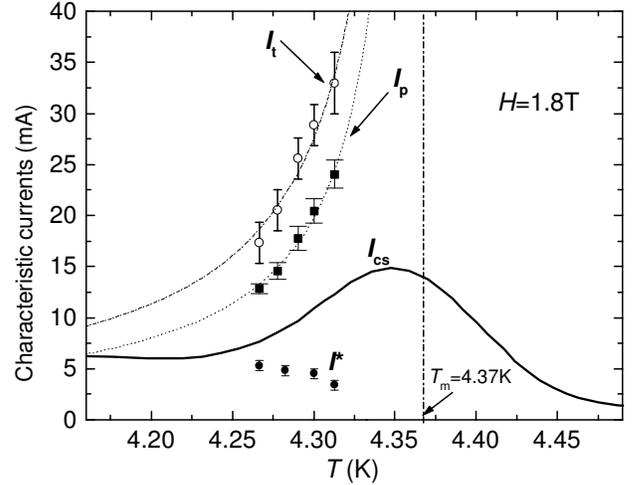} 
\protect\caption{ 
Temperature dependence of the currents separating the different dynamic regimes: 
$I<I^*$ - pinned; $I^*<I<I_{cs}$ - equilibration; $I_c<I<I_p$ - plastic flow; 
$I_p<I<I_t$ - smectic ordering; $I>I_t$ - crystalization. The dashed an dotted curves  show fits 
to the K-V model discussed in the text. }
\label{fig:fig_pl1}
\end{figure}

In order to identify the KV  crystallization we follow the usual approach in 
transport measurements\cite{bhat,kap}  focusing  on
the shape of the  differential resistance curve  shown in Fig. 3(b). The 
differential resistance is vanishly small for low currents and becomes finite 
only above $I_{cs}$ where metastability has disappeared. 
As the current is increased  further the curve rises to a maximum value of 
21.7 $m\Omega $ reached at $I_p=20.4mA$ and then drops down saturating at
the Bardeen-Stephen free flux flow value  $R_{BS}$= 18m$\Omega = R_n
H/H_{c2} $ at $I_t=$28.8mA. Its shape can be interpreted  according to
numerical simulations on the motional organization of a vortex
lattice\cite{kosh,shi,ols,kolt} :  at low velocities the motion is plastic
leading to  a defective lattice. The  defect density increases with
velocity resulting in a corresponding increase in differential resistance
which peaks at $I_p$. At this point  the  vortices  move in  an array of
almost periodic channels forming a smectic state characterized by
transverse order alone.  As the velocity is further increased the
vortices order inside the channels eventually  crystallizing at  $I_t$. 
The temperature dependence of  $I_t$, shown in Fig.4, 
is consistent with the KV prediction for the crystallization with  $T_m=4.37 \pm 
0.01$K. In the same figure we also show the curves for 
$I^*$,  $I_{cs}$, $I_p$ and   $I_t$ which separate the different 
 dynamic regimes of the moving vortex system as described in the caption.

In both sets of measurements presented here the  FC state is initially 
metastable and can reorganize into a stable state when driven by a current. The 
reorganization sets in at a current $I^*$ which is significantly lower  than the 
critical current of the pristine FC state. Below the peak region, the FC state 
undergoes an 
equilibration transition at $I^*$ which leads to an  ordered stable state 
indistinguishable from the ZFC state. By contrast the ZFC state is unaffected by 
the current. For the data in the peak effect region, vortex 
reorganization sets in at $I^*$ for $both$ the FC and ZFC states, one becoming 
more ordered the other more disordered until, at $I_{cs}$, they merge into a 
single state and loose the  distinction of their initial preparation. Thus, by analogy with the equilibration 
transition seen below the peak region, we  conclude  that in this part of 
the phase diagram the initial FC and ZFC states are both metastable and $I^*$ is 
the onset of a current driven equilibration transition to a mixed stable state.  

Several aspects of this data  can be 
understood in the context of recent Hall 
probe measurements on Fe doped samples\cite{palt} which have shown that 
in the peak effect region a stable disordered phase can be injected through 
a surface barrier at the sample edges. This mechanism can readily account for 
the disordering of the ZFC 
vortex lattice as indicated in Fig. 3 and in Ref. \cite{xiao}. In fact 
it is possible that $I^*$ is associated with a surface barrier\cite{palt1}, so that bulk   
vortices are not subject to a Lorentz force until $I^*$ is exceeded. Other features of the data 
can   be described  in terms of  the phenomenological parameter for 
 the distance that the disordered phase 
penetrates into the 
sample, $L_r(T,H,I)$.  For example $L_r\simeq 0$ below the peak where the ordered 
state  is stable ,  but it is finite  at low velocities in the  peak region where the two phases 
coexist.  At high velocities, where the random potential is averaged out in 
the moving vortex frame, $L_r\simeq 0$ which leads to a stable ordered state.

In summary, the experiments described here demonstrate that a metastable vortex 
lattice can undergo two types of current driven transitions. Below the peak 
effect region we observed a current driven equilibration transition from a 
metastable  disordered state to a  stable ordered one.   In the peak effect 
region we detected both the equilibration and the KV 
transitions. 

We thank A. Rosch, E. Zeldov and Y. Paltiel for 
useful discussions. Work supported
by  NSF-DMR-9705389 and DOE DE-FG02-99ER45742.


\begin{references}
\bibitem{com} In  the transport data of Ref. \cite{bhat} the transition is seen in the lower part of 
the peak effect region and disappears below the peak effect where the  SANS 
experiments are carried out.
\end{references}
\end{document}